\documentclass[aps,pra,reprint,showpacs,superscriptaddress]{revtex4-1} 
\usepackage{graphicx} 
\usepackage{booktabs}   
\usepackage{amsmath}    
\usepackage{amssymb}    
\usepackage{braket}     
\usepackage{graphicx}   
\usepackage{subcaption}
\usepackage[a4paper, scale=0.8]{geometry}     
\usepackage{titlesec}
\usepackage{xcolor}     
\usepackage[
  colorlinks=true,
  linkcolor=red,
  citecolor=blue,
  urlcolor=magenta
]{hyperref}
\usepackage{tikz}
\usetikzlibrary {arrows.meta}

\usepackage{graphicx}
\usepackage{amsmath,mathtools,verbatim}
\usepackage{amssymb}
\usepackage{mathtools}
\usepackage{amsmath}
\usepackage{bm} 
\usepackage{epstopdf}
\usepackage{braket}
\usepackage{float} 
\usepackage{placeins}   

\begin{document}
\title{\textbf{Preparing Quantum Backflow States by Large Momentum Transfer}}

\author{Yuchong Chen}\affiliation{Cavendish Laboratory, University of Cambridge,
JJ Thomson Avenue, Cambridge CB3 0US, United Kingdom}
\author{Yijun Tang}\affiliation{Cavendish Laboratory, University of Cambridge,
JJ Thomson Avenue, Cambridge CB3 0US, United Kingdom}

\begin{abstract}

Quantum backflow refers to the appearance of negative probability current in a state whose momentum distribution is essentially positive. We propose a scheme to prepare such states in a noninteracting Bose--Einstein condensate using large-momentum-transfer (LMT) atom interferometry. Our approach extends the single-pulse proposal of Palmero \cite{palmero2013detecting} by allowing one interferometer arm to undergo a tunable sequence of momentum-transfer pulses before recombination with a freely propagating arm. For realistic parameters for $^{88}\mathrm{Sr}$, the protocol generates interference states with tunable probability current and negligible negative-momentum contamination. We evaluate both the probability current and the critical-density criterion introduced by Palmero \cite{palmero2013detecting}, and identify parameter regimes in which the backflow signature is enhanced relative to the single-pulse scheme. These results present LMT interferometry as a flexible route for preparing candidate quantum-backflow states in cold-atom experiments.

\end{abstract}

\maketitle

\section{Introduction}

Quantum mechanics permits the counterintuitive possibility that a freely propagating state can exhibit negative probability current even though its momentum distribution is essentially positive. This phenomenon is known as standard quantum backflow. It was first identified in the quantum arrival-time problem by Allcock \cite{allcock1969time}, and later studied in detail by Bracken and Melloy \cite{bracken1994probability}. Recent work has refined the quantitative bound for standard quantum backflow, showing that the maximal probability transfer in the positive-momentum setting is approximately $c \approx 0.0384506$ \cite{fewster2025repeated}. Over the past decades, quantum backflow has been investigated in a variety of settings, including operator-based approaches \cite{penz2005new}, random wave functions \cite{berry2010quantum}, decay processes \cite{van2019decay}, ring geometries \cite{strange2012large,goussev2021quantum,goussev2024quantum}, higher dimensions \cite{barbier2023unbounded,paccoia2020angular}, open quantum systems \cite{mousavi2020dissipative,mousavi2020quantum,erratumdissipative}, scattering in the presence of potentials and defects \cite{bostelmann2017quantum,de2021quantum}, many-particle systems \cite{barbier2020quantum}, the effects of bosonic and fermionic statistics \cite{barbier2025quantum}, and periodic lattices \cite{goussev2025searching}.

Despite this long-standing theoretical interest, quantum backflow has not yet been observed experimentally. A particularly promising route is provided by density measurements in ultracold atomic Bose--Einstein condensates \cite{palmero2013detecting,mardonov2014interference}. In particular, Palmero \textit{et al.} derived a relation between probability current and particle density for two interfering wave packets with a relative phase linear in position \cite{palmero2013detecting}. In that framework, there exists a critical density $\rho_{\mathrm{crit}}$ such that whenever the measured density $\rho(x,t)$ falls below $\rho_{\mathrm{crit}}(x,t)$, the local probability current is necessarily negative. This result is especially important because it converts the current-based notion of backflow into a density criterion that is, in principle, accessible to fluorescence or absorption imaging. Related experimentally motivated formulations that relax the strict positive-momentum constraint have also been discussed in Refs.~\cite{barbier2021experiment,miller2021experiment}. Although quantum backflow itself has not yet been observed, classical optical analogs have recently been demonstrated \cite{eliezer2020observation,daniel2022demonstrating,ghosh2023azimuthal,zhang2025observation}.

More recently, the scope of experimentally relevant backflow studies has been broadened beyond the standard positive-momentum setting. In particular, Paterek and Goussev introduced a general formulation of quantum backflow for realistic wave packets with arbitrary momentum distributions, defining the genuinely nonclassical contribution as the excess probability flow beyond that expected from the corresponding classical momentum distribution alone \cite{paterek2026general}. Within that broader framework, the maximal general-backflow signal can reach nearly $13\%$, i.e.\ more than three times the standard Bracken--Melloy value for unidirectional states \cite{paterek2026general,fewster2025repeated}. The present work does not analyze the generalized backflow formulation in detail. Instead, we focus on the standard quantum backflow setting, where the momentum distribution is essentially positive.

In this work, we ask whether the cold-atom strategy of Palmero \textit{et al.} \cite{palmero2013detecting} can be extended to a large-momentum-transfer (LMT) atom-interferometric setting \cite{dimopoulos2008general}. Compared with a single-pulse preparation protocol, LMT interferometry offers additional control over the relative amplitudes of the two arms and over their final momentum separation before recombination. These quantities directly influence the interference pattern, the probability current, and the associated critical-density threshold. Our goal is therefore not to derive a new universal bound on quantum backflow, but rather to develop a more flexible preparation protocol for standard quantum backflow states.

Specifically, we consider a noninteracting $^{88}\mathrm{Sr}$ Bose--Einstein condensate in which one interferometer arm evolves freely while the other undergoes a tunable sequence of momentum-transfer pulses. We derive the final interfering state produced by this sequence, evaluate its probability current and critical density, and identify parameter regimes in which the backflow signature is enhanced relative to the single-pulse scheme of Ref.~\cite{palmero2013detecting} while negative-momentum contamination remains negligible. We also quantify the resulting backflow signal and discuss the accompanying experimental tradeoff: stronger backflow is associated with shorter density-modulation length scales, which may make direct imaging more demanding.

This paper is organized as follows. In Sec.~II we introduce the interferometric protocol and derive the final two-arm state. In Sec.~III we present numerical results for the probability current, the critical-density criterion, and the dependence of the backflow signal on the beam-splitter parameters. Section~IV concludes.

\section{Theory}
\label{section: Theory}

\begin{figure}[t]
    \centering
    \includegraphics[width=0.52\textwidth]{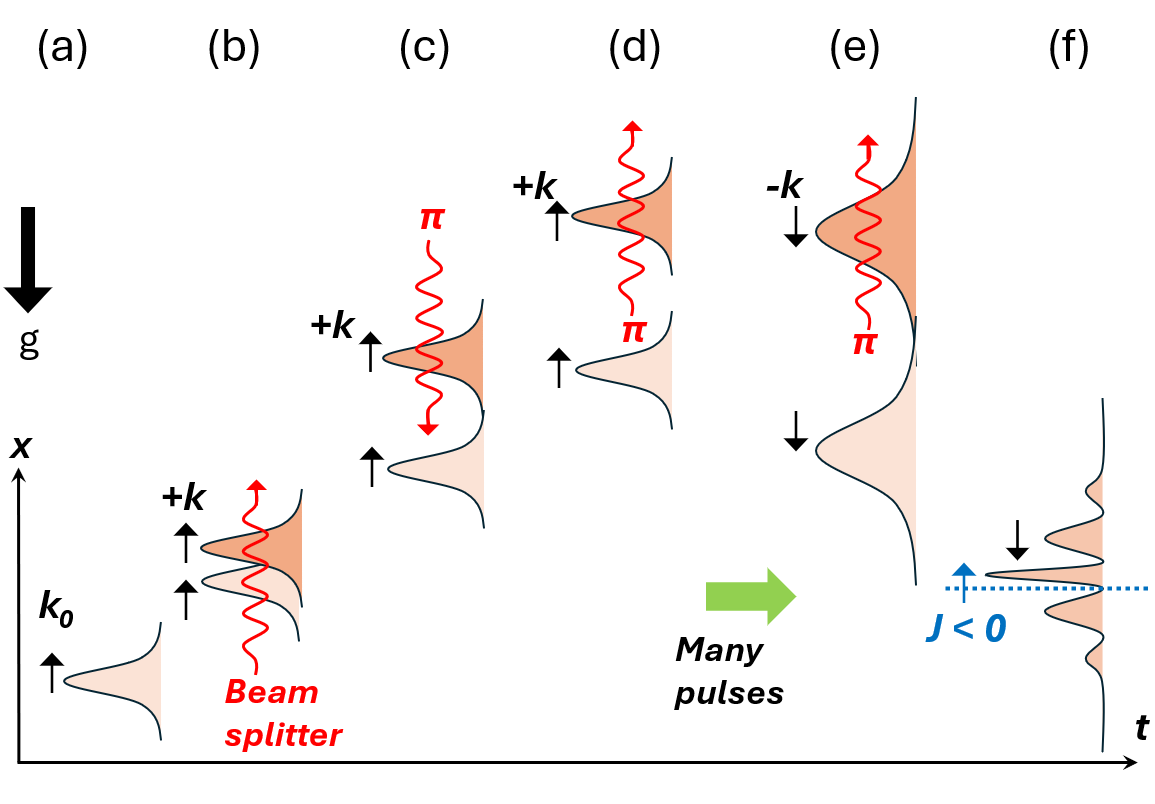}
    \caption{Schematic of the large-momentum-transfer interferometric protocol used to prepare candidate quantum-backflow states. (a) A Bose--Einstein condensate is launched upward with momentum $\hbar k_0$. (b) A beam-splitter pulse, not necessarily a $\pi/2$ pulse, creates two interferometer arms. (c) A $\pi$ pulse drives stimulated emission on one arm, imparting a recoil momentum $+\hbar k$, where $k$ is the laser wave number. (d) A subsequent $\pi$ pulse drives stimulated absorption, changing the internal state and imparting the corresponding recoil. (e) After many such pulses, the addressed arm acquires the desired momentum before recombination. (f) The two arms later overlap and interfere to form the final state.}
    \label{Schematics}
\end{figure}

As shown in Fig.~\ref{Schematics}, we begin with a cloud of cold two-level atoms in a Bose--Einstein condensate confined in a dipole trap and launched upward, as in a standard atom-interferometric sequence. Shortly after launch, a splitting pulse is applied to the condensate, creating two interferometer arms. The total state can then be written as the coherent superposition
\[
\ket{\Psi}=c_f\ket{\Psi_f}+c_b\ket{\Psi_b},
\]
where \(\ket{\Psi_f}\) and \(\ket{\Psi_b}\) denote the two arms after the splitter. The arm \(\ket{\Psi_f}\) remains in the ground internal state and subsequently undergoes free fall, while the arm \(\ket{\Psi_b}\) is transferred to the excited internal state and is later addressed by a large-momentum-transfer (LMT) pulse sequence, i.e., a sequence of light pulses designed to impart multiple photon recoils and thereby produce a larger momentum separation between the two interferometer arms.

\begin{figure}[tb]
    \centering
    \includegraphics[width=0.34\textwidth]{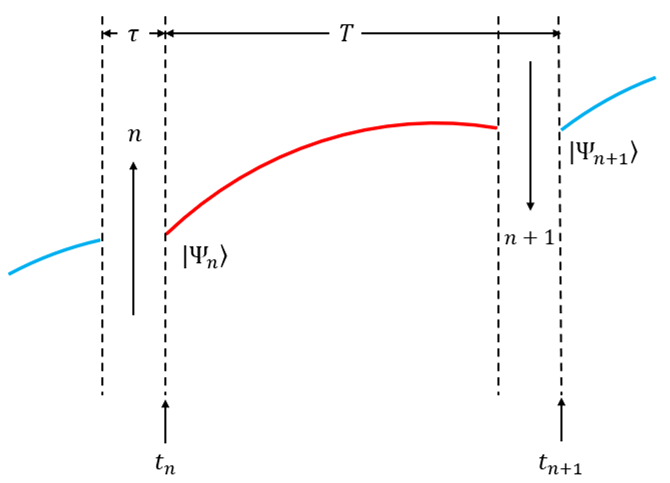}
    \caption{Pulse-to-pulse notation used for the LMT sequence. The state of the addressed arm immediately after the $n$th pulse is denoted by $\ket{\Psi_n}$. Each pulse has duration $\tau$, and adjacent pulses are separated by an interval $T=t_{n+1}-t_n$. After the next pulse, the state is denoted by $\ket{\Psi_{n+1}}$. Blue and red segments denote the ground and excited internal states, respectively.}
    \label{fig:lmt_pulse_timing}
\end{figure}

The LMT sequence consists of multiple laser pulses, generally applied in alternating directions. As the atomic wave packet passes through this pulse array in time, the pulses drive stimulated absorption and emission processes, as illustrated schematically in Fig.~\ref{fig:lmt_pulse_timing}. Each pulse can therefore change both the internal state and the momentum of the addressed atoms. The amount of transferred population is controlled by the pulse area, or equivalently by the pulse duration. In particular, a $\pi$ pulse transfers the entire addressed population to the other internal state, whereas a $\pi/2$ pulse creates an equal superposition of the two internal states.

After the splitting pulse, only one arm is subjected to the subsequent LMT sequence, so that its momentum is tailored by a controlled series of recoils while the other arm evolves freely. The two arms later overlap and interfere, thereby forming the final candidate backflow state. By tuning the pulse sequence after the initial splitter, one can control the relative arm amplitudes and the final momentum separation, and hence tune both the probability current and the associated critical density. In this way, the present protocol extends the single-pulse scheme of Ref.~\cite{palmero2013detecting} within a more flexible interferometric setting.

The theoretical description of this protocol consists of three parts: free evolution of the condensate between laser pulses, state modification during each pulse, and evaluation of the final probability current and critical-density criterion after recombination. We discuss these in Secs.~II A--II C.

\raggedbottom

\subsection{Free propagation of the BEC state}

The free evolution of the wave packet between laser pulses can be described in a moving frame by a Galilean frame-transformation operator $\hat G$ \cite{dimopoulos2008general}:
\begin{equation}
\begin{aligned}
\label{eq:propagation}
\ket{\Psi}
&= \hat G \ket{\phi_{\rm c}(t)} \otimes \ket{A_i(t)} \\
&= {\rm exp}\left(\frac{i}{\hbar}\int \mathcal{L}_c \, {\rm d}t \right)
   {\rm exp}\left(-\frac{i}{\hbar}\hat{p}\cdot \vec{x}_c \right) \\
&\quad \times
   {\rm exp}\left(\frac{i}{\hbar}\vec{p}_c \cdot \hat{x} \right)
   \ket{\phi_{\rm c}(t)} \otimes \ket{A_i(t)} .
\end{aligned}
\end{equation}

Here, the Lagrangian $\mathcal{L}_c$ describes the action accumulated by the center-of-mass motion, while $\vec x_{c}$ and $\vec p_{c}$ denote the center-of-mass position and momentum, respectively. The operator $\hat p \cdot \vec x_c$ generates the spatial translation, and $\vec p_c \cdot \hat x$ gives the momentum boost associated with the transformation to the moving frame. The remaining factor in Eq.~\eqref{eq:propagation} is the tensor product of the BEC center-of-mass state $\ket{\phi_{\rm c}(t)}$ and the internal atomic state $\ket{A_i(t)}$. The corresponding spatial wave function is therefore
\begin{equation}
\begin{aligned}
\label{eq: general state}
    \braket{\vec x|\Psi}
    = e^{\frac{i}{\hbar}\int \mathcal{L}_c \, {\rm d}t}
      e^{\frac{i}{\hbar}\vec p_c \cdot (\vec x-\vec x_c)}
      \braket{\vec x-\vec x_c | \phi_{\rm c}(t)}
      \otimes \ket{A_i(t)} .
\end{aligned}
\end{equation}

To obtain the center-of-mass-frame wave function $\braket{\vec x-\vec x_c|\phi_{\rm c}(t)}$, we begin with the noninteracting BEC wave function in the laboratory frame \cite{palmero2013detecting}:
\begin{equation}
\begin{aligned}
    \psi(x,t)
    &= \frac{1}{\sqrt{b}} \,
       \psi_0\left(\frac{x-v_1 t}{b}\right) \\
    &\quad \times
    {\rm exp}\left[
    i\frac{m}{2\hbar}x^2\frac{\dot b}{b}
    + i k_1 x \left(1-\frac{\dot b}{b}t\right)
    + i\beta
    \right] .
\end{aligned}
\end{equation}
Here,
\[
\psi_0(x)=\frac{1}{\pi^{1/4}\sqrt{a_x}}
{\rm exp}\left(-\frac{x^2}{2a_x^2}\right)
\]
is the initial wave function in the dipole trap, given by the ground state of a harmonic potential. The quantity $a_x=\sqrt{\hbar/(m\omega_x)}$ is the harmonic-oscillator length, $\omega_x$ is the trap frequency, $\beta$ is an irrelevant global phase, and $\hbar k_1 = m v_1$ is the condensate momentum. The factor
\[
b=\sqrt{1+\omega_x^2 t^2}
\]
describes the cloud expansion.

In the center-of-mass frame, the velocity-dependent terms vanish, and the wave function reduces to
\begin{equation}
\label{eq: cm wfc}
    \braket{x-x_c|\phi_{\rm c}(t)}
    =
    \frac{1}{\sqrt{b}}
    \psi_0\left(\frac{x-x_c}{b}\right)
    {\rm exp}\left[
    i\frac{m\dot b (x-x_c)^2}{2\hbar b}
    \right] .
\end{equation}

Physically, Eq.~\eqref{eq: cm wfc} describes a BEC wave packet localized at the center-of-mass position $x_c$ and expanding with scale factor $b$. Because this form contains a small negative-momentum component, it can give rise to classical backflow, which is distinct from the quantum backflow studied here.

In the present setup, the atoms are launched vertically, so the action accumulated between two pulses separated by a time interval $\Delta t$ is given by free-fall motion:
\begin{equation}
\label{Eq: free fall action}
    \Delta S_C
    =
    \left(\frac{P_{c,{\rm n}}^2}{2m}-mgx_{c,{\rm n}}\right)\Delta t
    - P_{c,{\rm n}} g \Delta t^2
    + \frac{1}{3} m g^2 \Delta t^3 ,
\end{equation}
where $P_{c,{\rm n}}$ and $x_{c,{\rm n}}$ are the center-of-mass momentum and position immediately after the $n$th pulse. This completes the description of the free evolution between pulses.

\subsection{Pulse influence}

We now consider the effect of a laser pulse on the atomic cloud. The general treatment follows Ref.~\cite{dimopoulos2008general}. We write the state $\ket{\Psi}$ as a superposition of components associated with the two internal atomic states, corresponding to the ground and excited levels:
\begin{equation}
    \ket{\Psi_{\rm total}} = \int {\rm d}p \sum_{i=1}^{2} c_i(p)\ket{p}.
\end{equation}
Here, $c_1(p)$ and $c_2(p)$ denote the momentum-space amplitudes for the ground and excited internal states, respectively. In the short-pulse limit, the laser-driven transition is described by
\begin{equation}
    \label{eq: pulse transition matrix}
    \begin{pmatrix}
        c_1'(p) \\
        c_2'(p+\hbar k)
    \end{pmatrix}
    =
    \begin{pmatrix}
        \Lambda_c & -i\Lambda_s e^{-i\phi_L} \\
        -i\Lambda_s^* e^{i\phi_L} & \Lambda_c
    \end{pmatrix}
    \begin{pmatrix}
        c_1(p) \\
        c_2(p+\hbar k)
    \end{pmatrix},
\end{equation}
where $\Lambda_c=\cos(|\Omega|\tau/2)$ and $\Lambda_s=\frac{\Omega}{|\Omega|}\sin(|\Omega|\tau/2)$. Here, $\Omega$ is the Rabi frequency such that $\Omega/|\Omega|=1$ as in Ref.~\cite{dimopoulos2008general}, $k$ is the laser wave number, $\phi_L$ is the laser phase, and $\tau$ is the pulse duration. For a $\pi$ pulse, i.e.\ when $|\Omega|\tau=\pi$, this reduces to
\begin{equation}
    \label{eq:pi_pulse_transition_matrix}
    \begin{pmatrix}
        c_1'(p) \\
        c_2'(p+\hbar k)
    \end{pmatrix}
    =
    \begin{pmatrix}
        0 & -ie^{-i\phi_L} \\
        -ie^{i\phi_L} & 0
    \end{pmatrix}
    \begin{pmatrix}
        c_1(p) \\
        c_2(p+\hbar k)
    \end{pmatrix}.
\end{equation}

In our simulations, the initial BEC wave packet from the dipole trap is first split into two arms by a laser pulse that is not necessarily a $\pi/2$ pulse. This produces two arms with different amplitudes, and these weights ultimately affect the backflow signal. We treat the evolution of each arm separately. For example, applying a $\pi$ pulse to an arm initially in the ground state $\ket{\Psi}$ gives
\begin{equation}
\begin{aligned}
\label{eq:in_phase_transition}
    \ket{\Psi'}
    &= \int {\rm d}p\, c_2'(p)\ket{p} = -ie^{i\phi_L}\int {\rm d}p\, c_1(p-\hbar k)\ket{p} \\
    &= -ie^{i\phi_L} e^{ik\hat{x}} \ket{\Psi}.
\end{aligned}
\end{equation}

This corresponds to absorption of a photon by an atom in the ground state, transferring it to the excited state and imparting a recoil momentum in the laser-propagation direction. To generalize this to either internal state and either pulse direction, we introduce an index $\mu$, with $\mu=1$ for the ground state and $\mu=-1$ for the excited state. The wave function immediately after the $(n+1)$th pulse can then be written in terms of the state after the $n$th pulse as

\begin{equation}
\begin{aligned}
    \label{eq: pulse-pulse evolution}
    \braket{x|\Psi'}
    &= -ie^{i\mu \phi_L}
    e^{\frac{i}{\hbar}\int \mathcal{L}_c{\rm d}t}
    e^{i\mu kx_c}
    e^{\frac{i}{\hbar}(p+\mu \hbar k)(x-x_c)} \\
    &\quad \times \braket{x-x_c|\phi_c(t)} \otimes \ket{A_i(t)} .
\end{aligned}
\end{equation}

Compared with the pre-pulse state $\braket{x|\Psi}$ in Eq.~\eqref{eq: general state}, this expression contains the laser phase $\phi_L$, an additional momentum boost $(p+\mu\hbar k)$, and a global phase shift $\mu kx_c$.

\subsection{Probability Flux and Critical Density}

We now trace the evolution of the two interferometer arms separately. The free arm undergoes pure free fall and is not affected by the subsequent laser pulses. Its state at the encounter time $T_f$ is obtained by substituting the free-fall action, the corresponding momentum shift, the center-of-mass wave function, and the internal-state evolution into Eq.~\eqref{eq: general state}. For clarity, the derivation is given explicitly in Appendix~A. The resulting spatial wave function is
\begin{widetext}
\begin{equation}
\begin{aligned}
    \label{eq: free arm wfc}
    \braket{x|\Psi_f}
    &= \frac{1}{(1+\omega_x^2T_f^2)^{1/4}\pi^{1/4}\sqrt{a_x}}
    {\rm exp}\left[\frac{i}{\hbar}\left(\frac{1}{2}mv_0^2T_f-mv_0gT_f^2+\frac{1}{3}mg^2T_f^3 \right) \right]
    {\rm exp}\left[\frac{i}{\hbar}m(v_0-gT_f)(x-x_c)\right] \\
    &\quad \times
    {\rm exp}\left[-\frac{(x-x_c)^2}{2a_x^2(1+\omega_x^2T_f^2)} \right]
    {\rm exp}\left[i\frac{m}{2\hbar}(x-x_c)^2\frac{\omega_x^2T_f}{1+\omega_x^2T_f^2}\right]
    \otimes {\rm exp}\left( -\frac{i}{\hbar}E_{\lambda}T_f \right)\ket{\lambda} .
\end{aligned}
\end{equation}
\end{widetext}
Here $E_\lambda$ is the internal-state energy of the free arm.

For the momentum-transferred arm, the state changes at each pulse because of both free propagation between pulses and laser-induced momentum transfer at the pulse times. We therefore construct the evolution iteratively. Specifically, we assume that the state immediately after the $n$th pulse is known. We then propagate this state freely over the interval $\Delta t=t_{n+1}-t_n$ using Eq.~\eqref{eq: general state}, and finally apply the $(n+1)$th $\pi$ pulse using Eq.~\eqref{eq:pi_pulse_transition_matrix} and Eq.~\eqref{eq:in_phase_transition}. In this way, Eq.~\eqref{eq: bounded arm iterative} gives the state immediately after the $(n+1)$th pulse in terms of the state immediately after the $n$th pulse. The detailed derivation is presented in Appendix~B. The result is
\begin{widetext}
\begin{equation}
\begin{aligned}
    \label{eq: bounded arm iterative}
    \braket{x|\Psi_{n+1}}
    &= -i{\rm exp}\left(i\mu_n \phi_L\right)
    {\rm exp}\left(i\mu_n k_{n+1}x_{n+1}\right)
    {\rm exp}\left[\frac{i}{\hbar}\left(\frac{1}{2} mv_{c, n}^2 \Delta t-mgx_{c, n}\Delta t-mv_{c, n}g\Delta t^2+\frac{1}{3}mg^2\Delta t^3 \right) \right] \\
    &\quad \times
    {\rm exp}\left[\frac{i}{\hbar}mv_{c, n+1}(x-x_c)\right]
    \frac{1}{(1+\omega_x^2t_{n+1}^2)^{1/4}\pi^{1/4}\sqrt{a_x}}
    {\rm exp}\left[-\frac{(x-x_c)^2}{2a_x^2 (1+\omega_x^2t_{n+1}^2)} \right] \\
    &\quad \times
    {\rm exp}\left[i\frac{m}{2\hbar}(x-x_c)^2\frac{\omega_x^2t_{n+1}}{1+\omega_x^2 t_{n+1}^2} \right]
    \tilde{\phi}_{L, {\rm tot}}\,
    \tilde{\psi}_{i, {\rm tot}}\,
    \tilde{\Psi}_{S}
    \otimes {\rm exp}\left(-\frac{i}{\hbar}E_{\mu_n}\Delta t \right)\ket{\mu_{n+1}} .
\end{aligned}
\end{equation}
\end{widetext}
Here the subscript $n$ labels quantities evaluated immediately after the $n$th pulse. The overall factor $-i\exp(i\mu_n\phi_L)\exp(i\mu_n k_{n+1}x_{n+1})$ arises from the action of the $(n+1)$th $\pi$ pulse. The exponential containing $v_{c,n}$ and $x_{c,n}$ represents the free-fall action accumulated during the interval $\Delta t$ between the $n$th and $(n+1)$th pulses. The factor $\exp[i m v_{c,n+1}(x-x_c)/\hbar]$ gives the updated momentum boost after the pulse, while the Gaussian envelope and quadratic phase are the center-of-mass wave function of Eq.~\eqref{eq: cm wfc} evaluated at time $t_{n+1}$. Finally, $\tilde{\phi}_{L,{\rm tot}}$, $\tilde{\psi}_{i,{\rm tot}}$, and $\tilde{\Psi}_{S}$ denote the laser, internal-state, and action phases accumulated during all earlier stages of the sequence, and the factor $\exp(-iE_{\mu_n}\Delta t/\hbar)\ket{\mu_{n+1}}$ accounts for the internal-state evolution over the interval together with the state transfer induced by the $(n+1)$th pulse.\\

For later convenience, we rewrite Eqs.~\eqref{eq: free arm wfc} and \eqref{eq: bounded arm iterative} in terms of overall phase factors and position-dependent envelopes. We also note that the two arms do not recombine immediately after the final pulse, but instead propagate for an additional time interval $\Delta T_f$ after the last pulse at time $T_N$. Their wave functions at the encounter time can therefore be written as
\begin{equation}
\begin{aligned}
    \label{eq: encountering free arm wfc}
    \braket{x|\Psi_f}
    &= \braket{x-x_c|\phi_{\rm c}(T_f)}\,{\rm exp}\!\left[i\theta_f(E_{\lambda},v_0,T_f)\right] \\
    &\quad \times {\rm exp}\!\left[\frac{i}{\hbar}m(v_0-gT_f)(x-x_c)\right],
\end{aligned}
\end{equation}
and
\begin{equation}
\begin{aligned}
    \label{eq: encountering bounded arm wfc}
    \braket{x|\Psi_b}
    &= \braket{x-x_c|\phi_{\rm c}(T_f)}\,{\rm exp}\!\left[i\theta_b(\{E_{\mu_n}\},\{t_n\},\{k_n\})\right] \\
    &\quad \times {\rm exp}\!\left[\frac{i}{\hbar}m(v_N-g\Delta T_f)(x-x_c)\right],
\end{aligned}
\end{equation}
where $T_f=T_N+\Delta T_f$ is the encounter time. The total wave function is then
\begin{equation}
\begin{aligned}
\label{eq:total_wfc}
\braket{x|\Psi}
&= \braket{x|\Psi_f}\Bigl[
c_f + c_b \exp\!\left(i\theta_b-i\theta_f\right) \\
&\qquad \times  \exp\!\left(
\frac{i}{\hbar}m(v_N+gT_N-v_0)(x-x_c)
\right)
\Bigr].
\end{aligned}
\end{equation}

Equation~\eqref{eq:total_wfc} has the same mathematical structure as the wave function considered by Palmero \textit{et al.}~\cite{palmero2013detecting}, namely an envelope multiplied by the interference factor of two components with a relative phase linear in position. Writing
\[
\braket{x|\Psi_f}=R(x,t)e^{i\theta(x,t)},
\]
we can therefore apply the density-based backflow criterion of Ref.~\cite{palmero2013detecting} directly to the present case. In that framework, backflow is guaranteed whenever the measured density $\rho(x,t)$ falls below a critical threshold $\rho_{\rm crit}(x,t)$. Since the atomic density can in principle be extracted from fluorescence imaging or absorption imaging, this criterion converts the current-based notion of backflow into an experimentally accessible density condition.

The corresponding critical density is
\begin{equation}
\begin{aligned}
    \label{eq: critical density}
    \rho_{\rm crit}
    = \frac{q}{q+2\partial_x\theta(x,t)}\,|R(x,t)|^2\bigl(|c_f|^2-|c_b|^2\bigr),
\end{aligned}
\end{equation}
where
\begin{equation}
    q=\frac{m}{\hbar}(v_N+gT_N-v_0)
\end{equation}
is the effective wave-number difference between the two arms, and
\begin{equation}
\begin{aligned}
    \label{eq: detailed expression for R}
    R(x,t)
    &= \frac{1}{(1+\omega_x^2T_f^2)^{1/4}}\frac{1}{\pi^{1/4}\sqrt{a_x}} \\
    &\quad \times {\rm exp}\left[-\frac{(x-x_c)^2}{2a_x^2}\frac{1}{1+\omega_x^2T_f^2}\right],
\end{aligned}
\end{equation}
\begin{equation}
\begin{aligned}
    \label{eq: detailed expression for theta}
    \theta(x,t)
    &= \frac{1}{\hbar}\left(\frac{1}{2}mv_0^2T_f-mv_0gT_f^2+\frac{1}{3}mg^2T_f^3-E_{\lambda}T_f\right) \\
    &\quad + \frac{1}{\hbar}\left[m(v_0-gT_f)(x-x_c)+\frac{m}{2}\frac{\omega_x^2T_f}{1+\omega_x^2T_f^2}(x-x_c)^2\right].
\end{aligned}
\end{equation}
These are simply the modulus and phase extracted from Eq.~\eqref{eq: free arm wfc}. The probability current then follows as
\begin{equation}
\begin{aligned}
    \label{eq: flux}
    \frac{m}{\hbar}J(x,t)
    &= \partial_x\theta\,|\Psi|^2 + q|R|^2|c_b|^2 \\
    &\quad + q|R|^2\,{\rm Re}\!\left(c_f^*c_b e^{iqx+i\theta_b-i\theta_f}\right),
\end{aligned}
\end{equation}
where we have generalized the expression to allow for complex coefficients $c_f$ and $c_b$.

\section{Simulation Results\label{section: Simulation Result}}
\subsection{Setup\label{subsection: setup}}

In the simulations, we consider a hypothetical noninteracting $^{88}\mathrm{Sr}$ Bose--Einstein condensate and use its $^1S_0\text{--}{}^3P_1$ transition for the LMT sequence, with laser wavelength $689\,\mathrm{nm}$. The ${}^3P_1$ excited state has a lifetime of $21.6\,\mathrm{\mu s}$, which makes it suitable for LMT. Large-momentum-transfer interferometry up to $141\hbar k$ has already been achieved in $^{88}\mathrm{Sr}$ atom interferometers \cite{rudolph2020large}.

\begin{figure*}
	\centering
	\includegraphics[width=0.95\textwidth]{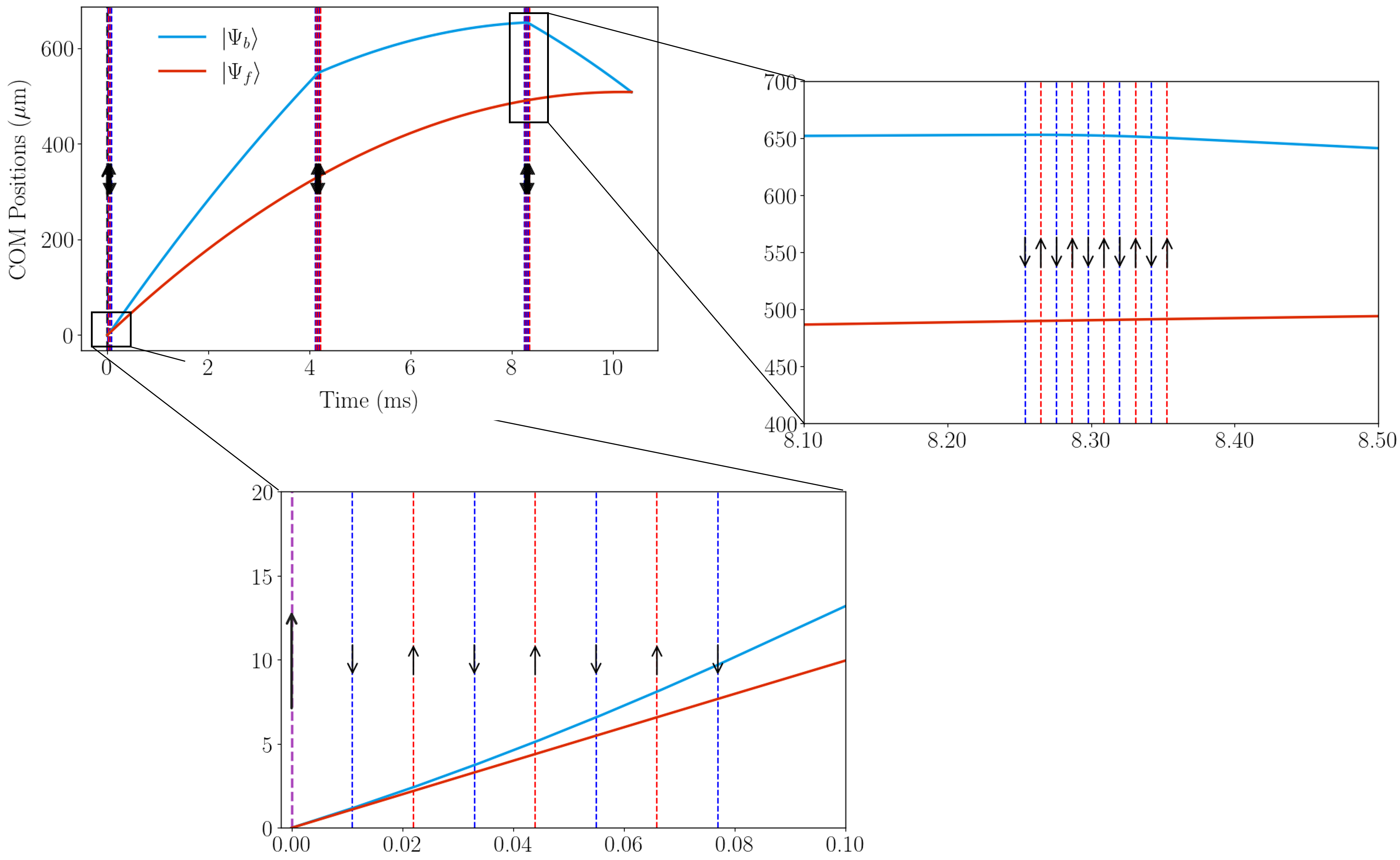}
	\caption{Simulation setup for preparing backflow states. The pulses address only $\ket{\Psi_b}$ (blue curve), so this arm undergoes large momentum transfer, while $\ket{\Psi_f}$ (red curve) evolves freely. At $t=0$, a splitting pulse (shown by the purple dashed line) transfers a fraction of the condensate into the excited state and creates the initial velocity difference between the two arms. The arm $\ket{\Psi_b}$ is then accelerated by a sequence of $\pi$ pulses.}
    \label{fig: setup}
\end{figure*}

We consider a BEC cloud launched upward with initial velocity $v_0=0.2\,\mathrm{m\,s^{-1}}$ and an initial spatial profile determined by a dipole trap with trap frequency $\omega_x=2\pi\times70\,\mathrm{s^{-1}}$ \cite{stellmer2013degenerate}. A beam-splitter pulse of tunable duration $\tau$ first splits the atoms into two arms with coefficients $c_b$ and $c_f$, according to Eq.~\eqref{eq: pulse transition matrix}. The subsequent geometry used in the simulation is shown in Fig.~\ref{fig: setup}.

Because of the narrow linewidth of the transition and the Doppler shift, only one arm is resonantly addressed by the later laser pulses. We refer to the unaddressed arm as the free arm and to the addressed arm as the pulsed arm. The free arm undergoes free fall without further laser interaction until it recombines with the pulsed arm. The pulsed arm is initially transferred to the excited state by the beam-splitter pulse and then undergoes 7 LMT pulses at wavelength $689\,\mathrm{nm}$, each with $\pi$-pulse duration, so that the population is transferred sequentially between the two internal states while acquiring momentum recoils. After this, the pulsed arm continues to propagate upward for $4\,\mathrm{ms}$ before being redirected downward by two pulse arrays, which are themselves separated by $4\,\mathrm{ms}$. Adjacent pulses within the sequence are separated by $11\,\mathrm{\mu s}$. The sequence is chosen so that, at the encounter point, the free arm has as small a downward velocity as possible, thereby maximizing the momentum difference between the two arms.

\subsection{Classical Backflow}

\begin{figure}
    \centering
    \includegraphics[width=0.45\textwidth]{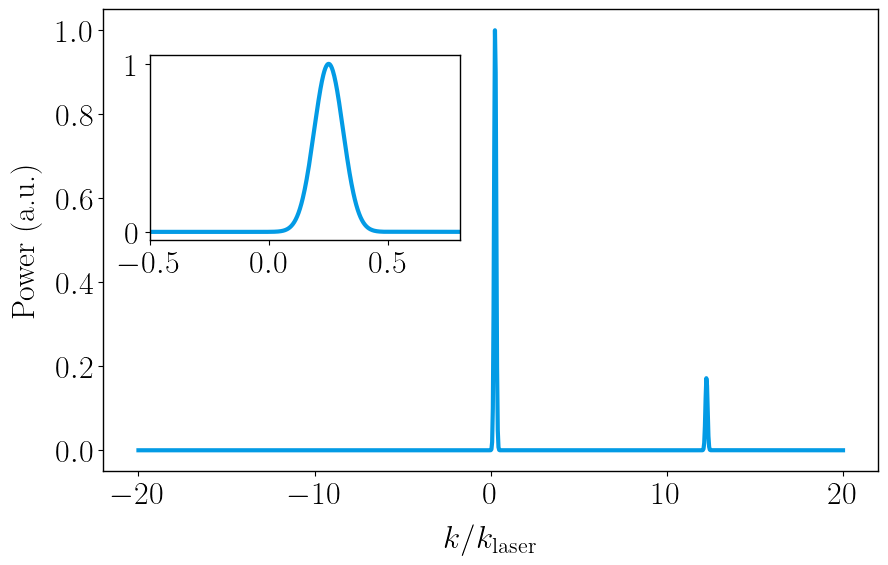}
    \caption{Momentum-space distribution of the final combined state. The two dominant peaks correspond to the momenta of the two interferometer arms at the encounter time. The negligible weight at negative momentum confirms that classical backflow contamination is suppressed in the parameter regime considered here.}
    \label{fig: momentum spectrum}
\end{figure}

We first verify that the final state contains negligible negative-momentum contribution, so that any observed negative current can be attributed to standard quantum backflow rather than to classical backflow arising from wave-packet expansion. In the present setting, classical backflow would originate from the small negative-momentum tail naturally generated as the Gaussian condensate expands during propagation. Following Ref.~\cite{palmero2013detecting}, this contribution can be made negligible by choosing parameters such that $mv/\hbar \gg 1/a_x$ and $R_0 < v/\omega_x$, where $R_0$ is the initial width of the wave packet. To confirm that this condition is satisfied in our simulations, we transform $\braket{x|\Psi}$ to momentum space and inspect the resulting distribution, shown in Fig.~\ref{fig: momentum spectrum}. The spectrum exhibits negligible support at negative momentum, indicating that the backflow signal studied here is not dominated by classical contamination.

For completeness, we note that recent work has introduced a more general formulation of quantum backflow for realistic wave packets, in which the classical contribution associated with the momentum distribution is subtracted explicitly from the total probability flow \cite{paterek2026general}. The present work does not analyze that generalized framework in detail; rather, we focus on the standard positive-momentum setting adopted in Ref.~\cite{palmero2013detecting}. Nevertheless, the LMT preparation strategy developed here may provide a useful starting point for future studies in that broader setting.

\subsection{Probability Flux}
\begin{figure*}
    \hspace{-0.5cm}
    \subfloat[$J$ vs.\ $x-x_c$\label{fig: Flux vs. Position 0.6pi}]
    {
        \begin{minipage}{8cm}
            \centering
            \includegraphics[scale=0.35]{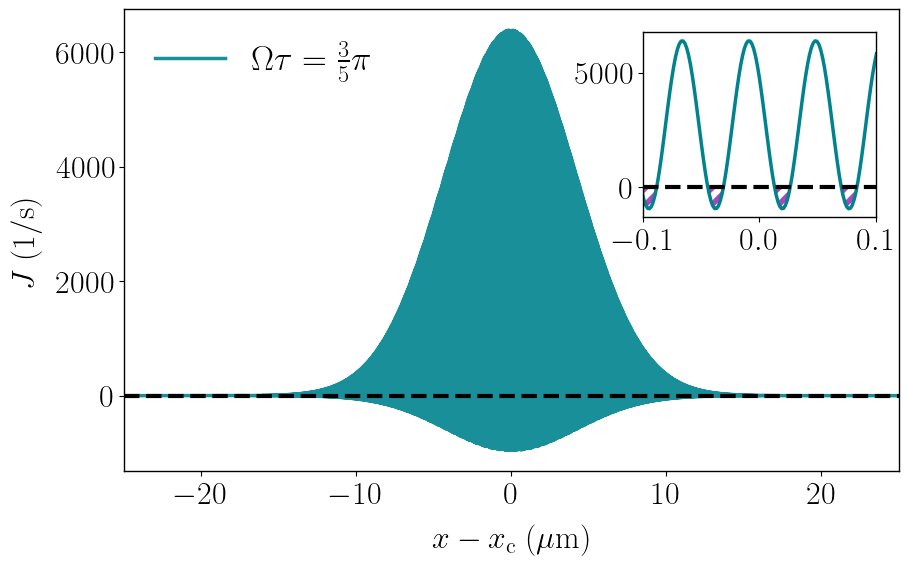}
        \end{minipage}
    }
    \subfloat[$|\Psi|^2$ vs.\ $x-x_c$\label{fig: critical density 0.6pi}]
    {
        \begin{minipage}{8cm}
            \centering
            \includegraphics[scale=0.35]{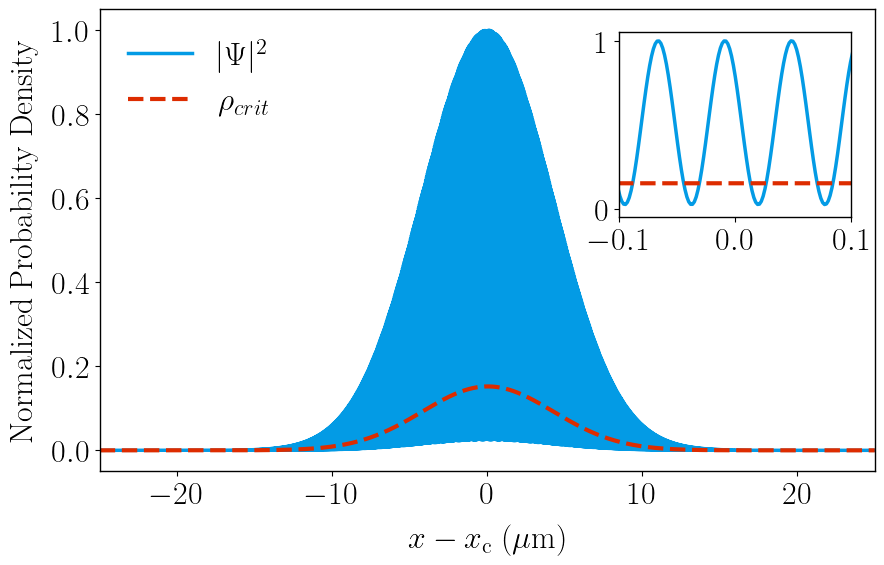}
        \end{minipage}
    }
    \caption{Example of backflow for an initial splitting parameter $\Omega\tau=0.6\pi$. (a) Probability flux $J$ as a function of position: a finite portion of the flux lies below zero, indicating the presence of backflow. (b) Normalized density profile $|\Psi|^2$ together with the critical density threshold: a corresponding portion of the density lies below $\rho_{\rm crit}$. At the center of the condensate, $\rho_{\rm crit}-\rho$ is $15.17\%$ of the maximum density.}
    \label{fig: Flux vs. Position 0.9pi}
\end{figure*}

The total wave function at the encounter point is used to evaluate the probability flux in its vicinity. Figure~\ref{fig: Flux vs. Position 0.6pi} shows an example of the spatial backflow profile for an initial splitting pulse with $\Omega\tau=0.6\pi$. Near the center of mass, approximately one sixth of the flux is negative. Another prominent feature of the flux distribution is its rapid oscillation in space. This follows from Eq.~\eqref{eq: flux}, in which $J$ contains an interference term oscillating with wave number $q$, set by the momentum difference between the two arms. In the present case, the velocity difference at the encounter point is $0.079\,\mathrm{m\,s^{-1}}$, which leads to the observed rapid spatial oscillations.

In Fig.~\ref{fig: critical density 0.6pi}, we plot the density profile of $\ket{\Psi}$ near the center of mass in blue, together with the critical density $\rho_{\rm crit}$ in red. Both curves are normalized by the maximum value of $|\Psi|^2$. The density profile of the final state consists of a Gaussian envelope modulated by oscillations. The Gaussian envelope originates from the initial ground-state wave function in the dipole trap, while the oscillatory structure arises from the position-dependent phase factor in Eq.~\eqref{eq:total_wfc}. From Fig.~\ref{fig: critical density 0.6pi}, the maximum value of $\rho_{\rm crit}$ is $15.21\%$, while the minimum of $|\Psi|^2$ near $x_c$ is $0.04\%$.

\subsection{Parameter Tuning}

We now show that the backflow signature can be tuned by varying the duration $\tau$ of the initial beam-splitter pulse. According to Eq.~\eqref{eq: pulse transition matrix}, this is equivalent to varying the Rabi phase $\Omega\tau$ experienced by the initial condensate, and therefore to tuning the arm coefficients $c_b$ and $c_f$.

In the experimental setting considered here, the condensate enters the splitter pulse entirely in the ground state. We therefore take $c_1(p)=1$ and $c_2(p)=0$ in Eq.~\eqref{eq: pulse transition matrix}, which gives
\[
c_b=\cos(\Omega\tau/2), \qquad c_f=-i\sin(\Omega\tau/2).
\]
Thus, the coefficients produced directly by the splitter pulse are, in general, complex. These complex coefficients are the ones most directly connected to the experimental preparation protocol studied in the present work. For each choice of $\Omega\tau$, we evaluate the probability flux using the fixed velocities and LMT sequence described in Sec.~\ref{subsection: setup}.

To compare different parameter settings, we define a spatial backflow measure as the area of the negative part of the flux profile. This quantity characterizes the spatial extent and magnitude of the negative-current region for a fixed encounter configuration.

\begin{figure*}[t]
    \centering
    \subfloat[Complex coefficients\label{fig: backflow rate complex}]
    {
        \includegraphics[width=0.47\textwidth]{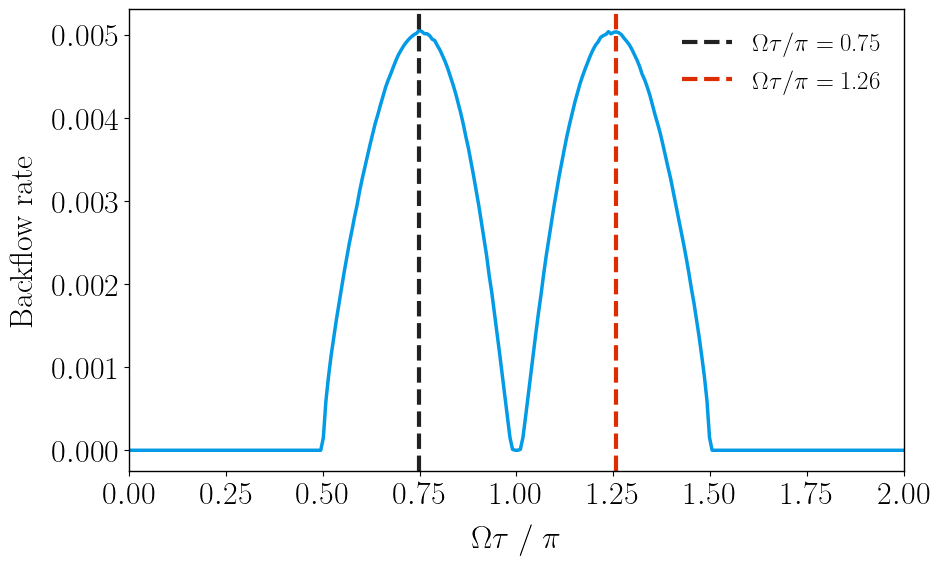}
    }
    \hfill
    \subfloat[Real coefficients\label{fig: backflow rate vs. cb}]
    {
        \includegraphics[width=0.47\textwidth]{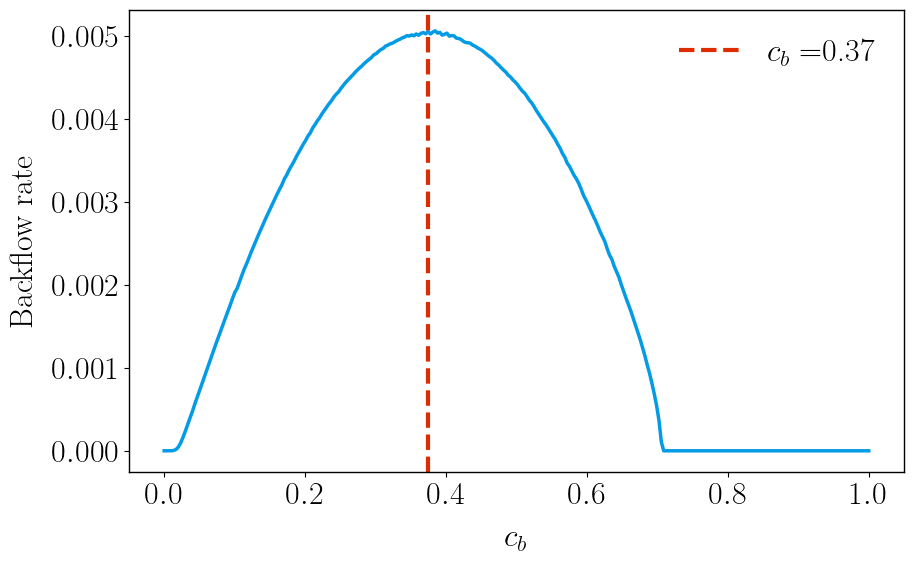}
    }
    \caption{Dependence of the spatial backflow measure on the arm coefficients. (a) The measure is plotted against the splitter phase $\Omega\tau$ for the coefficients generated directly by the beam-splitter pulse, which are generally complex. Two backflow windows appear symmetrically about $\Omega\tau=\pi$. (b) For comparison, the same measure is plotted against $c_b$ in an auxiliary scan with real coefficients. In this case a single peak is obtained. For $c_b>1/\sqrt{2}$, $\rho_{\rm crit}$ becomes negative and no backflow occurs.}
    \label{fig: Flux rate}
\end{figure*}

Figure~\ref{fig: backflow rate complex} shows this measure as a function of $\Omega\tau$ for the experimentally generated complex coefficients. The curve is symmetric about $\Omega\tau=\pi$, with one backflow peak on each side. There are also intervals, $\Omega\tau\in[0,\pi/2]\cup[3\pi/2,2\pi]$, in which no backflow occurs. These correspond to parameter values for which $|c_f|^2<|c_b|^2$, so that Eq.~\eqref{eq: critical density} gives a negative $\rho_{\rm crit}$. The dip at $\Omega\tau=\pi$ corresponds to a $\pi$ pulse, which transfers the entire initial ground-state population into the excited arm and therefore produces no interference backflow.

\begin{figure}[H]
    \centering
    \subfloat[$J$ vs.\ $x-x_c$\label{fig: Flux vs. Position 0.75pi}]
    {
        \includegraphics[width=0.47\textwidth]{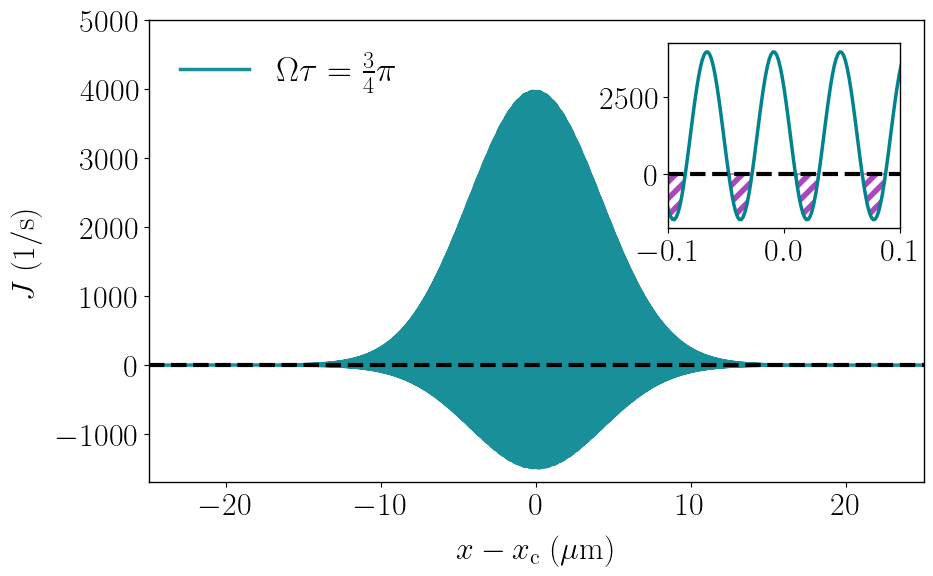}
    }
    \hfill
    \subfloat[$|\Psi|^2$ vs.\ $x-x_c$\label{fig: critical density 0.75pi}]
    {
        \includegraphics[width=0.47\textwidth]{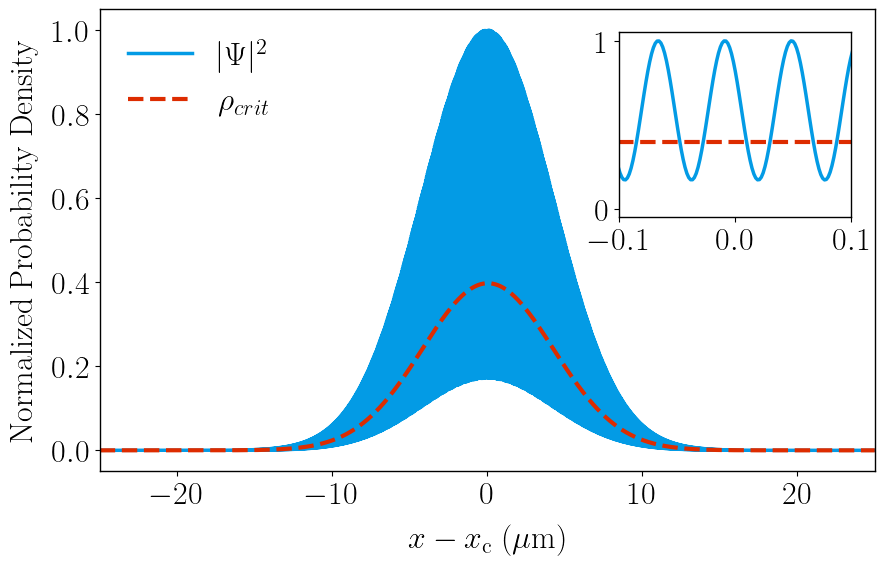}
    }
    \caption{Backflow at the parameter value $\Omega\tau=0.75\pi$, where the spatial backflow measure is maximal. (a) Probability flux $J$: approximately one third of the peak flux is negative. (b) Normalized density profile $|\Psi|^2$: at the center of the cloud, $\rho_{\rm crit}-\rho$ reaches $21.40\%$ of the maximum density.}
    \label{fig: Flux vs. Position 0.75pi_all}
\end{figure}

We next focus on the parameter value $\Omega\tau=0.75\pi$, which maximizes the spatial backflow measure in Fig.~\ref{fig: backflow rate complex}. The corresponding flux and density profiles are shown in Fig.~\ref{fig: Flux vs. Position 0.75pi}. The maximum negative flux is below $-1000\,\mathrm{s^{-1}}$, and approximately one third of the flux profile is negative. Comparing Figs.~\ref{fig: critical density 0.6pi} and \ref{fig: critical density 0.75pi}, we find that the region where $\rho<\rho_{\rm crit}$ remains of similar spatial extent, approximately $\pm20\,\mathrm{\mu m}$, because the overall cloud size is mainly set by the trap length scale $a_x$ and the propagation time, which are unchanged across the simulations. At the same time, the critical-density threshold itself is substantially enhanced: the minimum of $|\Psi|^2$ near $x_c$ is $18.39\%$, while the maximum $\rho_{\rm crit}$ reaches $39.79\%$, more than twice the value reported in the single-pulse proposal of Ref.~\cite{palmero2013detecting}, where $\rho_{\rm crit}\approx17\%$.

For the experimentally relevant splitter considered here, the coefficients $c_b$ and $c_f$ are generally complex, as discussed above. For completeness, and to facilitate comparison with the earlier idealized treatment of Ref.~[1], we also perform an auxiliary scan in which $c_b$ and $c_f$ are taken to be real while satisfying $c_b^2+c_f^2=1$. This real-coefficient case does not correspond to the bare splitter output alone; rather, it represents a phase-engineered variant in which an additional controllable phase shift is applied after the splitter and before the LMT sequence. As shown in Fig.~\ref{fig: backflow rate vs. cb}, this produces a single peak, and no backflow occurs in the limiting cases where only one arm is present and no interference is possible. The maximum value of the spatial backflow measure is comparable to that obtained for complex coefficients. Backflow again disappears for $c_b>1/\sqrt{2}$, where $c_f<c_b$ and $\rho_{\rm crit}$ becomes negative. The maximum occurs at $c_b=0.37$. Compared with the complex-coefficient case, the real-coefficient scan produces a broader parameter window in which backflow is present.

Finally, to connect our results with the standard integrated backflow probability transfer discussed in Refs.~\cite{bracken1994probability,fewster2025repeated,paterek2026general}, we evaluate the flux $J$ at the encounter location over a short time interval before the two arms overlap fully. Integrating the negative part of this time-dependent flux yields an integrated backflow probability transfer of $\Delta=0.003514$. This is the quantity that should be compared with the Bracken--Melloy bound. We stress that our claim of enhancement in the present work refers instead to the density-based signature, in particular the larger critical-density threshold relative to Ref.~\cite{palmero2013detecting}, rather than to any claim of exceeding the known universal bound on integrated backflow probability transfer.
\begin{figure}[H]
    \centering
    \includegraphics[width=0.92\columnwidth]{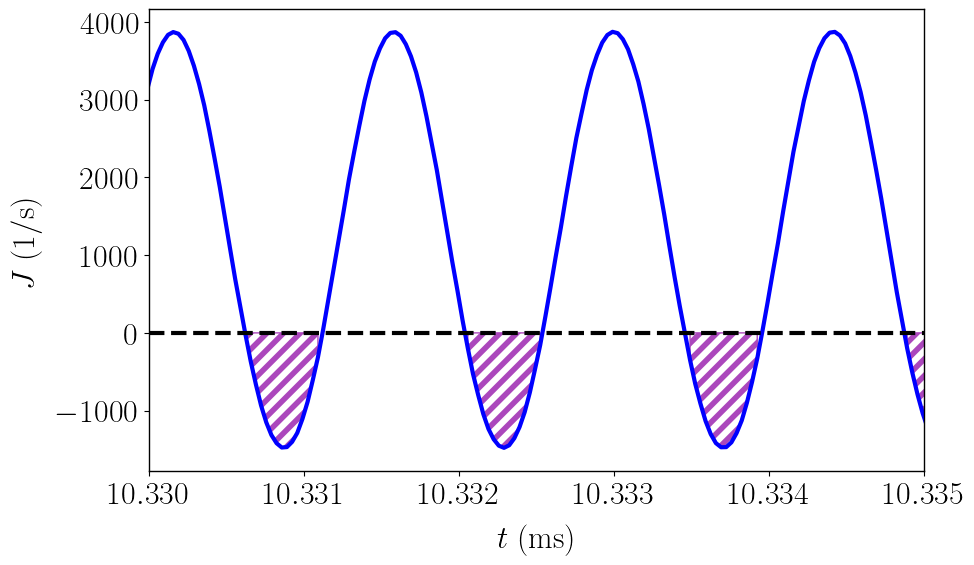}
    \caption{Time-dependent flux $J$ at a position near the encounter point. Although the magnitude of the negative flux is relatively large, the time interval over which $J<0$ is short, giving an integrated backflow probability transfer of $\Delta=0.003514$.}
    \label{fig: flux vs. time}
\end{figure}

\section{Conclusion}
\label{section: Conclusion}

To prepare quantum-backflow states in a more flexible manner, we have proposed a protocol that extends the single-pulse cold-atom scheme of Palmero \textit{et al.}~\cite{palmero2013detecting} to a large-momentum-transfer atom-interferometric setting. In contrast to Ref.~\cite{palmero2013detecting}, the present approach provides additional control over the relative arm amplitudes and the final momentum separation through the initial beam-splitting pulse and the subsequent sequence of $\pi$ pulses applied to one interferometer arm. Using realistic experimental parameters, we have shown that this protocol produces final interference states with tunable probability current and negligible negative-momentum contamination.

We have also analyzed the associated critical-density criterion and identified a practical tradeoff: stronger backflow is accompanied by a shorter density-modulation length scale, which makes conventional fluorescence imaging more challenging. The main outcome of the present work is therefore a more versatile interferometric framework for preparing and diagnosing candidate quantum-backflow states in a cold-atom platform. It would be interesting in future work to examine how the same LMT-based preparation strategy may be adapted to the more general-backflow framework for realistic wave packets discussed in Ref.~\cite{paterek2026general}. Since that framework is formulated in terms of position measurements at multiple distinct times, it is plausible that fluorescence or absorption imaging of the atomic density at different evolution times could provide a useful route toward implementing the required repeated position measurements in a cold-atom platform.

\appendix
\onecolumngrid
\section*{Appendix}
\renewcommand{\thesubsection}{\thesection\arabic{subsection}}
\setcounter{section}{0}
\section{Free Arm}

For the free arm, there is no influence from the subsequent LMT pulses. Its final state can therefore be derived analytically from Eq.~\eqref{eq: general state}. We evaluate each contribution in turn. Let $T_f$ denote the time at which the two arms recombine, $v_0$ the initial launch velocity of the BEC cloud, $P_c$ the center-of-mass momentum, and $m$ the atomic mass. The total action accumulated during free fall is then
\begin{equation}
\begin{aligned}
\label{detailed eq: free fall action}
   \Delta S
   = \int_0^{T_f} \mathcal L_c \, {\rm d}t
   = \int_0^{T_f}\left( \frac{P_c^2}{2m}-mgx_c \right) {\rm d}t
   = \frac{1}{2}mv_0^2T_f - mv_0gT_f^2+\frac{1}{3}mg^2T_f^3 .
\end{aligned}
\end{equation}
Here we assume that the cloud starts at $x_c(t=0)=0$.

The momentum-shift term in Eq.~\eqref{eq: general state} is
\begin{equation}
\label{detailed eq: momentum shift for free fall}
    \vec{P}_c\cdot(\vec x-\vec{x}_c)=m(v_0-gT_f)(x-x_c) .
\end{equation}

The BEC wave function in the center-of-mass frame is obtained by writing out Eq.~\eqref{eq: cm wfc} explicitly:
\begin{equation}
\label{detailed eq: cm wfc}
\braket{\vec x-\vec x_c|\phi_c(T_f)}
=
\frac{1}{(1+\omega_x^2T_f^2)^{1/4}}\frac{1}{\pi^{1/4}\sqrt{a_x}}
\exp\left[-\frac{(x-x_c)^2}{2a_x^2}\frac{1}{1+\omega_x^2T_f^2}\right]
\exp\left[i\frac{m}{2\hbar}(x-x_c)^2\frac{\omega_x^2T_f}{1+\omega_x^2T_f^2} \right] .
\end{equation}
The prefactor contains the expansion factor $1/\sqrt{b}$, while the first exponential is the ground-state harmonic-trap wave function modified by the free expansion.

For the internal-state evolution, the free arm remains in the same internal level throughout the sequence. Denoting its initial internal state by $\ket{\lambda}$ and the corresponding energy by $E_\lambda$, the time evolution is
\begin{equation}
\label{detailed eq: internal state free fall}
    \ket{A_i(T_f)}=\exp\left( -\frac{i}{\hbar}E_{\lambda}T_f \right)\ket{\lambda} .
\end{equation}

Substituting Eqs.~\eqref{detailed eq: free fall action}, \eqref{detailed eq: momentum shift for free fall}, \eqref{detailed eq: cm wfc}, and \eqref{detailed eq: internal state free fall} into Eq.~\eqref{eq: general state} gives the final state of the free arm:
\begin{equation}
\begin{aligned}
    \label{detailed eq: free arm iterative}
    \braket{x|\Psi_f}
    &= \frac{1}{(1+\omega_x^2T_f^2)^{1/4}\pi^{1/4}\sqrt{a_x}}
    {\rm exp}\left[\frac{i}{\hbar}\left(\frac{1}{2}mv_0^2T_f-mv_0gT_f^2+\frac{1}{3}mg^2T_f^3 \right) \right]
    {\rm exp}\left[\frac{i}{\hbar}m(v_0-gT_f)(x-x_c)\right] \\
    &\quad \times
    {\rm exp}\left[-\frac{(x-x_c)^2}{2a_x^2(1+\omega_x^2T_f^2)} \right]
    {\rm exp}\left[i\frac{m}{2\hbar}(x-x_c)^2\frac{\omega_x^2T_f}{1+\omega_x^2T_f^2}\right]
    \otimes {\rm exp}\left( -\frac{i}{\hbar}E_{\lambda}T_f \right)\ket{\lambda} .
\end{aligned}
\end{equation}

\section{Pulsed Arm}

The pulsed arm undergoes successive changes in momentum, phase, and internal state throughout the LMT sequence. For this reason, it is not convenient to write its final state in a single closed analytic form. Instead, we construct the evolution iteratively.

Specifically, we assume that the state immediately after the $n$th pulse is known, and then determine the state immediately after the $(n+1)$th pulse in two stages. First, the arm evolves freely during the interval $\Delta t=t_{n+1}-t_n$ according to Eq.~\eqref{eq: general state}. Second, at time $t_{n+1}$ it encounters the next laser pulse, which induces the corresponding phase shift, internal-state transfer, and momentum recoil according to Eq.~\eqref{eq:pi_pulse_transition_matrix} and Eq.~\eqref{eq:in_phase_transition}. Repeating this procedure pulse by pulse yields the final state at the end of the full LMT sequence.

Since the pulse duration $\tau$ is much shorter than the separation between adjacent pulses, $t_n-t_{n-1}$, we neglect the free-fall evolution during each laser interaction. The free propagation between pulses has the same general form as for the free arm. In particular, the momentum-shift term after the $(n+1)$th pulse is
\begin{equation}
    \label{detailed eq: momentum shift pulsed}
    \vec{P}_{c,n+1}\cdot(\vec x-\vec x_c)=mv_{c,n+1}(x-x_c),
\end{equation}
and the corresponding center-of-mass wave function at time $t_{n+1}$ is
\begin{equation}
    \label{detailed eq: cm wfc pulsed}
    \braket{x-x_c|\phi_c(t_{n+1})}
    =
    \frac{1}{(1+\omega_x^2t_{n+1}^2)^{1/4}}
    \frac{1}{\pi^{1/4}\sqrt{a_x}}
    \exp\left[
    -\frac{(x-x_c)^2}{2a_x^2}\frac{1}{1+\omega_x^2t_{n+1}^2}
    \right]
    \exp\left[
    i\frac{m}{2\hbar}(x-x_c)^2
    \frac{\omega_x^2t_{n+1}}{1+\omega_x^2t_{n+1}^2}
    \right].
\end{equation}

All pulses after the initial splitter are taken to be $\pi$ pulses. Since the pulsed arm starts in the excited state, it alternates between the excited and ground internal states throughout the sequence. We denote the internal state immediately after the $n$th pulse by $\ket{\mu_n}$, with $\mu_n=1$ for the ground state and $\mu_n=-1$ for the excited state. From Eq.~\eqref{eq:pi_pulse_transition_matrix} and Eq.~\eqref{eq:in_phase_transition}, the $(n+1)$th pulse imparts a momentum recoil $\Delta P_c=\mu_n\hbar k$ together with a global phase factor $-i\exp(i\mu_n\phi_L)\exp(i\mu_n kx_c)$. The corresponding internal-state evolution over the interval $\Delta t$ is
\begin{equation}
    \label{detailed eq: internal state pulsed}
    \ket{A_i(t_{n+1})}
    =
    \exp\left( -\frac{i}{\hbar}E_{\mu_n}\Delta t \right)\ket{\mu_{n+1}}.
\end{equation}

To write the full state after the $(n+1)$th pulse, we must also include the phases accumulated during all previous stages of the sequence. We denote these by $\tilde{\phi}_{L,{\rm tot}}$ for the accumulated laser phase, $\tilde{\Psi}_S$ for the accumulated action phase, and $\tilde{\psi}_{i,{\rm tot}}$ for the accumulated internal-state phase. Combining the free propagation over $\Delta t$ with the action of the $(n+1)$th pulse then gives
\begin{equation}
\begin{aligned}
    \label{detailed eq: bounded arm iterative}
    \braket{x|\Psi_{n+1}}
    &= -i{\rm exp}\left(i\mu_n \phi_L\right)
    {\rm exp}\left(i\mu_n k_{n+1}x_{n+1}\right)
    {\rm exp}\left[
    \frac{i}{\hbar}\left(
    \frac{1}{2}mv_{c,n}^2\Delta t
    -mgx_{c,n}\Delta t
    -mv_{c,n}g\Delta t^2
    +\frac{1}{3}mg^2\Delta t^3
    \right)
    \right] \\
    &\quad \times
    {\rm exp}\left[\frac{i}{\hbar}mv_{c,n+1}(x-x_c)\right]
    \frac{1}{(1+\omega_x^2t_{n+1}^2)^{1/4}\pi^{1/4}\sqrt{a_x}}
    {\rm exp}\left[-\frac{(x-x_c)^2}{2a_x^2(1+\omega_x^2t_{n+1}^2)}\right] \\
    &\quad \times
    {\rm exp}\left[
    i\frac{m}{2\hbar}(x-x_c)^2
    \frac{\omega_x^2t_{n+1}}{1+\omega_x^2t_{n+1}^2}
    \right]
    \tilde{\phi}_{L,{\rm tot}}\,
    \tilde{\Psi}_S
    \otimes
    \tilde{\psi}_{i,{\rm tot}}\,
    {\rm exp}\left(-\frac{i}{\hbar}E_{\mu_n}\Delta t\right)
    \ket{\mu_{n+1}}.
\end{aligned}
\end{equation}

In the simulations, we update the pulsed-arm state iteratively using Eq.~\eqref{detailed eq: bounded arm iterative}, and finally combine it with the free-arm state in Eq.~\eqref{detailed eq: free arm iterative} at the encounter time to obtain the probability flux and number density.

\twocolumngrid
\bibliographystyle{unsrt}
\bibliography{ref}

\end{document}